# The strange (hi)story of particles and waves[*]

H. Dieter Zeh – www.zeh-hd.de – arxiv:1304.1003v23

**Abstract:** This is a non-technical presentation (in historical context) of the quantum theory that is strictly based on global unitarity. While the first part is written for a general readership, Sect. 5 may appear a bit provocative. I argue that the single-particle wave functions of quantum mechanics have to be correctly interpreted as *field modes* that are "occupied once" (that is, first excited states of the corresponding quantum oscillators in the case of boson fields). Multiple excitations lead non-relativistically to apparent many-particle wave functions, while the quantum states proper are always defined by wave function(al)s on the configuration space of fundamental fields, or on another, as yet elusive, fundamental local basis.

**Contents:**

**Overview:** Sects. 1 and 2 are a brief review of the early history - neglecting details. Sects. 3 and 4 concentrate on some important properties of non-relativistic quantum mechanics that are often insufficiently pointed out in textbooks (including quite recent ones). Sect. 5 describes how this formalism would have to be generalized into its relativistic form (QFT), although this program generally fails in practice for *interacting* fields because of the complicated entanglement that would arise between too many degrees of freedom. This may explain why QFT is mostly *used* in a semi-phenomenological manner that is often misunderstood as a fundamentally new theory. Sect. 6 describes the application of this Schrödinger picture to quantum gravity and quantum cosmology, while Sect. 7 concludes the paper.

---

[*] Free and extended translation of my unpublished German text "Die sonderbare Geschichte von Teilchen und Wellen" – available at my website since October 2011. By the term "(hi)story" I tried to catch the double-meaning of the German word "Geschichte". V15 has now been published in Z. Naturf. A, **71**, 195 (2016).
– Depending on your interests, you may prefer to skip the first one or two (historical) sections!

# 1. Early History

The conceptual distinction between a discrete or a continuous structure of matter (and perhaps other "substances") goes back at least to the pre-Socratic philosophers. However, their concepts and early ideas were qualitative and speculative. They remained restricted to some general properties, such as symmetries, while the quantitative understanding of continuous matter and motion had to await the conceptual development of calculus on the one hand, and the availability of appropriate clocks on the other. Quantitative laws of nature and the concept of mass points, for example, were invented as part of classical mechanics.

This theory was first applied to extended "clumps of matter", such as the heavenly bodies or falling rocks and apples. It was in fact a great surprise for Newton and his contemporaries (about 1680) that such very different objects – or, more precisely, their centers of mass – obeyed the same laws of motion.[1] The objects themselves seemed to consist of continuous matter, although the formal concept of mass points was quite early also applied to the *structure* of matter, that is, in the sense of an atomism. Already in 1738, Daniel Bernoulli explained the pressure of a gas by the mean kinetic energy of presumed particles, but without recognizing its relation to the phenomenon of heat. If one regarded these particles themselves as small elastic spheres, however, the question for their internal structure would in principle arise anew. The concept of elementary particles thus appears problematic from the outset.

At about the same time, Newton's theory was also generalized by means of the concept of a continuum of infinitesimal mass points which can move according to their local interaction with (mainly their repulsion by) their direct neighbors. This route to continuum mechanics required novel mathematical concepts, but no fundamentally new *physical* ones beyond Newton. The assumption of an unlimited divisibility of matter thus led to a consistent theory. In particular, it allowed for wave-like propagating density oscillations, required to describe the phenomenon of sound. So it seemed that the fundamental question for the conceptual structure of matter had been answered.

As a byproduct of this "substantial" (or "Laplacean") picture of continuum mechanics, based on the assumption of distinguishable and individually moving infinitesimal elements of matter, also the elegant "local" (or "Eulerian") picture could be formulated. In the latter, one neglects any reference to trajectories of individual elements in order to consider only its spatial density distribution together with a corresponding current density as the kinematical objects of interest. In modern language they may be called a scalar and a vector *field*. In spite of this new form, however, continuum mechanics remained based on the concept of a locally conserved material substance consisting of individual elements.

This model for a continuum of mass points would be incomplete if the latter could move freely, interrupted only by occasional collisions, as suspected for a gas by Daniel Bernoulli. Since his gas pressure (which allows for sound waves, too) is given by the density of molecular kinetic energy, that is, by the product of the number density of gas particles and their mean kinetic energy, this could still be understood as representing a "chaotic continuum" by means of an appropriately defined simultaneous limit of infinite particle number density and vanishing particle size. This remained a possibility even when chemists began to succeed in applying Dalton's and Avogadro's hypotheses about molecular structures from the beginning of the nineteenth century in order to understand the chemical properties of the various substances. Similar to Auguste Bravais's concept of crystal lattices (about 1849), these structures were often regarded as no more than a heuristic tool to describe the internal structure of a multi-component continuum. This view was upheld by many even after Maxwell's and Boltzmann's explanation of thermodynamic phenomena in terms of molecular kinetics, and in spite of repeated but until then unsuccessful attempts to determine a finite value for Avogadro's or Loschmidt's numbers. The "energeticists", such as Wilhelm Ostwald, Ernst Mach and initially also Max Planck remained convinced until about 1900 that atoms are an illusion, while concepts like internal energy, heat and entropy would describe fundamental continua (fields). Indeed, even after the determination of Loschmidt's number could they have used an argument that formed a severe problem for atomists: Gibbs' paradox of the missing entropy of self-mixing of a gas. Today it is usually countered by referring to the *indistinguishability* of molecules of the same kind, although the argument requires more, namely the *identity* of states resulting from their permutations. Such an identity would be in conflict with the concept of particles with their individual trajectories, while a *field* with two bumps at points $x$ and $y$ would by definition be the *same* as one with bumps at $y$ and $x$ Although we are using quite novel theories today, such conceptual subtleties do remain essential – see Sect. 5. (Their role in statistical thermodynamics depends also on dynamical arguments.)

Another object affected by the early dispute about particles and waves is light. According to its potential of being absorbed and emitted, light was traditionally regarded as a "medium" rather than a substance. Nonetheless, and in spite of Huygens' early ideas of light as a wave phenomenon in analogy to sound, Newton tried to explain it by means of "particles of light", which were supposed to move along trajectories according to the local refractive index of matter. This proposal was later refuted by various interference experiments, in particular those of Thomas Young in 1802. It remained open, though, what substance (called the ether) did oscillate in space and time – even after light had been demonstrated by Heinrich Hertz in 1886 to represent an electromagnetic phenomenon in accordance with Maxwell's equations. The possibility of these fields to propagate and carry energy gave them a certain substantial character that seemed to support the world of continua as envisioned by the energeticists. Regarding atoms, Ernst Mach used to ask "Have you ever seen one?" whenever somebody mentioned them to him. Later in this article I will argue that his doubts may be justified even today – although we *seem* to observe individual atoms and particle tracks. Similar to the phenomenon of "events" or "quantum jumps", they may be an illusion caused by the dynamics of Schrödinger's wave function, which does *not* live in space (Sect. 3).

At the end of the nineteenth century, the continuum hypothesis suffered a number of decisive blows. In 1897, J. J. Thomson discovered the elementary electric charge; in 1900, Max Planck postulated his radiation quanta for the electromagnetic field with great success; and in 1905, Albert Einstein estimated the value of Loschmidt's number $N_L$ by means of his theory of Brownian motion. Thereafter, even the last energeticists resigned, but they left some confusion about the concept of physical "states". While they had regarded temperature, pressure and internal energy density etc. as local degrees of freedom that characterize individual (ontic) states of matter, the latter were in atomic physics replaced by "thermodynamic states" that average over unknown particle properties, either in time, or in space ("coarse graining"), or with respect to some other incomplete knowledge. In quantum theory, this confusion survives in the operationalist definition of states or in the concept of "mixed states" (see Sect. 4).

Einstein even revived the concept of particles of light (later called photons) – although he regarded it merely as a "heuristic point of view" that he was never ready to fully accept himself. For some time, Planck's radiation quanta were indeed attributed to a discrete emission process rather than to the radiation itself. So in 1913, Niels Bohr replaced the concept of classical motion for atomic electrons by stochastic "jumps" between his discrete atomic orbits – in accordance with Planck's and Einstein's ideas about a probabilistic radiation process. These early ideas led later to the insufficient interpretation of quantum mechanics as no more than stochastic dynamics for otherwise classical particles.

However, the development soon began to proceed in the opposite direction again.[2] In 1923, Louis de Broglie inverted Einstein's speculative step from light waves to photons by postulating a wave length $\lambda = c/\nu = h/p$ for the electron, where $p$ is its momentum, in analogy with Planck's relation $E = pc = h\nu$. For him, this could only mean that all microscopic objects must consist of both, a particle *and* a wave, whereby the wave has to serve as a "guiding field" or "pilot wave" for the particle. This field would have to be more powerful than a conventional *force* field, since it has to determine the velocity rather than merely the acceleration; the initial velocity can according to this proposal not be freely chosen any more once the wave function is given. When David Bohm later brought this theory into a consistent form, it turned out that the pilot wave cannot be defined in space ("locally"), since it has to be identified with the global entangled wave function to be described in Sect. 4.

## 2. Wave Mechanics

Inspired by de Broglie's ideas, Schrödinger based his novel wave mechanics of 1926 on the assumption that electrons are *solely* and uniquely described by wave functions (spatial fields, as he first thought). His wave equation allowed him to explain the hydrogen spectrum by replacing Bohr's specific electron orbits by standing waves. In this way he could explain the puzzling discrete quantum numbers by the numbers of nodes the wave function needs to obey its boundary conditions. For a special case (the harmonic oscillator) he was furthermore able to construct "wave packets" that may imitate *moving* particles – see Fig. 1 for the case of *free* motion, however. Shortly thereafter, interference phenomena in agreement with de Broglie's wave length were observed by Davisson and Germer for electrons scattered from crystal lattices. A wave function can furthermore penetrate a potential barrier and thus quantitatively explain "quantum tunneling" for the phenomenon of α-decay. Does this not very strongly indicate that electrons and other "particles" are in reality just wave packets of fields that obey Schrödinger's wave equation?

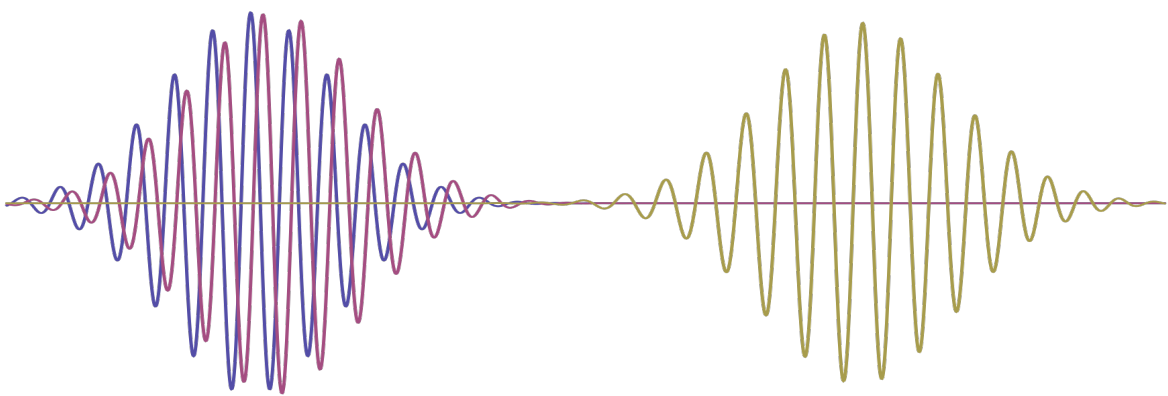

**Fig. 1:** Real part of a one-dimensional complex wave packet (the product of a Gaussian with a plane wave $e^{2\pi i x/\lambda}$) moving freely according to the time-dependent Schrödinger equation, depicted at three different times (blue: t=0,

red: t=0.04, yellow: t=1 in relative units). If the wave packet describes reality, its width defines a "real uncertainty" for the object position; it does neither represent incomplete information, nor is it related to the measureable "particle" size (which has to be described by *internal* degrees of freedom – see Sect. 4). When comparing blue and red, one recognizes that the packet moves faster than its wave crests, while the yellow curve demonstrates a slight spreading of the packet (in contrast to the mentioned harmonic oscillator). The center of the packet moves according to the group velocity $v = p/m := h/m\lambda$, where the mass $m$ is just a parameter of the wave equation. For this reason, momentum is in wave mechanics *defined* by the wave number $h/\lambda$ (not by motion!), although it is mostly *observed* by means of moving wave packets (moving "objects"). It can then be measured even for plane waves, which would not define a group velocity, by means of the conservation of wave numbers $k = 2\pi/\lambda$ during interactions with objects that do exist as wave packets, thus giving rise to the concept of "momentum transfer". Already for atomic masses and thermal velocities, the de Broglie wave length is clearly smaller than the radius of a hydrogen atom, so one may construct quite narrow wave packets for their center of mass (cms) wave functions. While the dispersion of the wave packet decreases with increasing mass $m$, it becomes always non-negligible after a sufficient time interval. In order to compensate for it, one would need an additional dynamical mechanism that permanently reduces the "coherence length" characterizing a wave packet in order to retain the appearance of a particle (see for "collapse" or "decoherence" in Sect. 4).

A few months before Schrödinger invented his wave mechanics, Heisenberg had already proposed his matrix mechanics. In contrast to Schrödinger, he did not abandon the concept of particles, but in a romantic attempt to revive Platonic idealism and overcome a mechanistic world view, combined with an ingenious guess, he introduced an abstract formalism that was to replace the concept of deterministic trajectories by formal probabilistic rules. Together with Born and Jordan, Heisenberg then constructed an elegant algebraic framework that could be used to "quantize" all mechanical systems. This mathematical abstraction perfectly matched Heisenberg's idealistic philosophy. In particular, matrix mechanics was shown *in principle* to lead to the same predictions as wave mechanics – although it could be used in practice only in simple cases. A year after his first paper, Heisenberg supplemented his formalism by his uncertainty relations between position and momentum of an electron or other "conjugate" pairs of variables. Such a fundamental uncertainty is clearly in conflict with a consistent concept of particles, while in wave mechanics it is simply a consequence of the Fourier theorem – without any uncertainty *of the wave function* or the assumption of an unavoidable "distortion" of the state of the electron during a measurement (as originally suggested by Heisenberg). Another indication of a choice of inappropriate concepts may be the requirement of a "new logic" for them. So it is not surprising that Schrödinger's intuitive wave mechanics was preferred by most atomic physicists – for a short time even by Heisenberg's mentor Max Born. For example, Arnold Sommerfeld wrote only a "Wellenmechanischer Ergänzungsband" to his influential book "Atombau und Spektrallinien".

Some important phenomena, though, remained in conflict with Schrödinger's theory. While his general wave equation $i\hbar\partial\psi/\partial t = H\psi$ would allow various time-dependent solutions, such as the moving wave packet of Fig. 1, bound electrons appeared to be restricted to standing waves. The latter are solutions of the stationary Schrödinger equation $H\psi = E\psi$ that gives rise to the observed discrete eigenvalues $E_n$ under the required boundary conditions. Although this equation can be derived from the general one under the assumption of a special time dependence of the form $\psi \propto e^{iEt/\hbar}$, there is no general reason for this special form. Instead of obeying the time-dependent equations when interacting with electromagnetic fields, these bound states seemed to be ruled by Bohr's stochastic "quantum jumps", which would thus explain energy quanta of radiation (including the hydrogen spectrum) by means of the conservation of energy. Other wave functions seem to "jump" or "collapse" into particle-like narrow wave packets during position measurements. In a Wilson chamber, one could even observe tracks of droplets that can be regarded as successions of such position measurements along particle trajectories.

As a consequence, Schrödinger seemed to resign when Max Born, influenced by Wolfgang Pauli, re-interpreted his new probability postulate, which originally was to describe jumps between different wave functions, in terms of probabilities for the spontaneous *creation of particle properties* (such as positions or momenta). This interpretation turned out to be very successful (and earned Born a Nobel prize) even though it was never quite honest, since the wave function does *not only* describe probabilities. It is also required to represent individual observable properties, such as energy or angular momentum, by means of corresponding "eigenstates", whose structure can often be confirmed by appropriate experiments. Similarly, a spinor (a generalized wave function for the electron spin) describes probabilities for other *individual* spinor states rather than for classical properties.

The impossibility to derive the successful wave function from his uncertainty principle (while the reverse *is* possible) was so painful for Heisenberg that he regarded the former as "a new form of human knowledge as an intermediary level of reality", while Bohr introduced his, in his own words, "irrational" principle of complementarity. It required the application of mutually exclusive ("complementary") classical concepts, such as particles and waves, to the same objects. No doubt – this was an ingenious pragmatic strategy to avoid many problems, but, from there on, the search for a unique description of Nature was not allowed any more in microscopic physics. Pure *Gedanken*-experiments, traditionally used as consistency tests for physical concepts, were now discredited for being "counterfactual". As an answer to the ques-tion whether the electron be *really* a wave or a particle (or what else), Bohr insisted that "there is no microscopic reality" – a conclusion that was often regarded as philosophically very deep. Only few dared to object that "this emperor is naked", while the term "complementarity" might be no more than another name for a conceptual inconsistency – thus perhaps forming the greatest sophism in the history of science. (A related misconception is the recently invented "quantum information" about quantities that cannot really *exist*.) The large number of attempts of a philosophical or formal explanation of this "nonconcept" is even the more impressive. Furthermore, the question when and where precisely the probability interpretation (or the "Heisenberg cut" between quantum and classical concepts) has to be applied, that is, when a "virtual" property becomes "real", remained open to be pragmatically decided from case to case. Therefore, the Hungarian Eugene Wigner spoke of a "Balkanization of physics" – a traditional (Hapsburgian) expression for the decay of law and order during that time.

## 3. Wave Functions in Configuration Space

So one should take a more complete look at Schrödinger's wave mechanics. When he formulated it, he used Hamilton's partial differential equations as a guiding principle. These equations, the result of a reformulation of classical mechanics, are solved by a scalar function whose gradient describes a continuum of independent classical trajectories which differ by their initial conditions – sort of a wave function without interference. Hamilton had mainly been interested in the elegant mathematical form of this theory rather than in applications. This turned out to be an advantage for Schrödinger. He assumed that Hamilton's equations were no more than a short wave lengths approximation (corresponding to the limit $h \to 0$) of a *fundamental* wave theory – similar to the approximation of geometric optics in Maxwell's theory. However, this short wave length approximation only means that local parts of an extended wave propagate almost independently of one another along spatial paths – not that they represent particles. Similarly, Feynman's path integral defines a propagating *wave* as a superposition of the various causal chains contained in such a continuum,[3] while it neither requires nor justifies the existence of individual paths that might then simply be selected by an increase of information. Different partial waves or Feynman paths can in fact interfere with one another (that is, they may have coherent physical effects). This means that they exist together as *one reality* (one wave function) rather than merely defining a statistical ensemble of *possibilities*. They could be turned into an ensemble only by some stochastic dynamics that would have partially to replace the deterministic wave equation.

While light waves propagate in three-dimensional space, Hamilton's waves must according to their construction exist in the configuration space of all possible classical states $q$ of the system under consideration. Therefore, Schrödinger, too, obtained wave functions on (what appears to us classically as) configuration spaces of various dimensions rather than in space. This is an enormous difference, that turned out to be very important for atoms and molecules. Intuitive *spatial* wave functions are here quite insufficient, in general. These "true" wave functions can also be understood as a consequence of Dirac's fundamental superposition principle, since the superposition of all classical configurations $q$ defines precisely a wave function $\psi(q)$ on configuration space. It can then easily be further generalized to include properties that never occur as classical variables (such as spin, neutrino flavor, or even the difference between a *K*-meson and its antiparticle), whose superpositions may again define new *individual* physical states (even new kinds of "particles"). Dirac himself understood his superpositions in Born's pragmatic but enigmatic sense as "probability amplitudes" for properties that are formally represented by Heisenberg's classically motivated "observables". There is no absolutely preferred basis in Hilbert space, and probabilities are thus meaningful only with respect to corresponding "measurements". If these observables are written in terms of dyadic products of their eigenstates (their spectral representation), they may formally describe Born's probabilities as those for *jumps of wave functions* (stochastic projections in Hilbert space as part of the dynamics). Any proposal for some fundamental theory underlying quantum mechanics would first of all have to explain the very general and well established superposition principle, which, in particular, describes all phenomena of quantum nonlocality without any "spooky" action at a distance (see Sect. 4). There does not seem to be any way to *derive* this principle and its corresponding probability interpretation from some classical field theory with its merely *spatial* superposition principle – no matter how many dimensions.

Schrödinger was still convinced of a reality in space and time, and so he tried, in spite of the Hamiltonian analogy, to understand the electron wave function as a spatial *field* (the "$\psi$-field"). Therefore, he first restricted himself with great success to single-particle problems (quantized mass points, whose configuration space is isomorphic to space). This approach misled not only himself for some time, but a whole generation of physicists. A spatial wave function can also be readily used to describe scattering problems – either in terms of the center-of-mass wave function of an object scattered from a potential, or of the relative coordinates of a two-body problem. In scattering events, Born's probability interpretation is particularly suggestive because of the usual subsequent position measurement in a detector. A wave function in space is indeed usually meant when one speaks of the *wave-particle dualism*. In spite of its limited and often misleading role, three-dimensional wave mechanics still dominates large parts of most textbooks because of its success in correctly and intuitively describing many important single-particle aspects, such as the energy spectrum of the hydrogen atom and scattering probabilities. It is often supported by presenting the two-slit experiment as *the* key to understand quantum mechanics, although this is only one specific aspect of the theory.

The generalization (or rather the return) to wave functions in configuration space happened almost unnoticed at those times of great confusion – for some physicists even until today. While most of them are now well aware of "quantum nonlocality", they remain used to arguing in terms of spatial waves for many purposes. In contrast to classical fields, however, even single-particle wave functions do not describe additive (extensive) quantities, since a piece cut from a plane wave representing a quantum "particle", for example, would describe its full charge and kinetic energy (the latter being defined by the wave number).

Schrödinger took initially great pains to disregard or to re-interpret his general wave equation in configuration space, even though it is precisely its application to oscillating field amplitudes instead of mass points that explains Planck's radiation quanta $h\nu$. (Another early example is the rigid rotator, whose wave function depends on the three Euler angles.) The reason is that the spectrum $E = nh\nu$ for quantum oscillators $q_i$ (here the amplitudes of given field modes that classically oscillate in time with different frequencies $\nu_i$ rather than positions of mechanical oscillators) is proportional to the natural numbers $n$. Only this specific spectral property describes additive *energy quanta $h\nu$* regardless of any emission process that had often been made responsible for the quantization of radiation. So it also explains the concept of "occupation numbers" for the field modes (or "photon" numbers). In Schrödinger's wave mechanics, these quantum numbers $n$ are explained by the number of nodes of the oscillator wave functions. These have to be distinguished from the spatial nodes of the field modes themselves (such as $sin(k_i x)$ multiplied by a polarization vector). These field modes (rather than their wave functions) then appear as "photon wave functions" – see below and Sect. 5.

But where can one find these oscillator wave functions if not in space? In contrast to Fig. 1, they are here defined as functions on the abstract configuration space of the field amplitude $q_i$. Different eigenmodes of a *classical* field $q(\mathbf{x},t)$, such as plane waves with their classical wave numbers $k_i = 2\pi\nu_i/c$, can fortunately be quantized separately; their Hamiltonians commute. This means that energy eigenstates $\Psi$ for the total quantum field factorize in the form $\Psi=\Pi_i\psi_i(q_i)$, while their eigenvalues simply add, $E = \Sigma_i E_i$. Although the oscillator spectrum $E_i = n_i h \nu_i$ can also be derived from Heisenberg's algebra of observables (matrix mechanics) without explicitly using wave functions, the latter's nodes for a fixed field mode $q_i$ have recently been made visible for various "photon number" eigenstates (similar to different energy eigenfunctions of the electron in the hydrogen atom) in an elegant experiment.[4] The wave functions $\psi_i(q_i)$ on configuration space have thus been demonstrated to "exist", although they cannot be attributed to the traditional wave-particle dualism, which would refer to *spatial* waves characterizing "quantum particles". The importance of this fundamental experiment for the wave-particle debate has in my opinion not yet been appropriately appreciated by the physics community or in textbooks (see Sect. 5 for further details).

The difference between Schrödinger's theory and a classical field theory becomes particularly obvious from the fact that the amplitudes of a classical field now appear as *arguments q* in Schrödinger's wave function. Positions occur here only as an "index" that distinguishes field amplitudes at different space points, where they form a spatial continuum of *coupled* oscillators. Since classical fields are usually written as functions on space and time, $q(x,t)$, the confusion of their spatial arguments with particle positions in the single-particle wave function $\psi(x,t)$ led to the questionable concept of a "time operator" in an attempt to restore space-time symmetry. However, $x$ and $t$ in the field $q$ are both classical coordinates, while the particle position $x$ in $\psi$ defines dynamical degrees of freedom (still called "variables" although they now appear only as arguments of the time-dependent wave function).

While a general time-dependent "one-photon wave function" can be understood as a quantum superposition of different modes of the electromagnetic field (its Fourier representation) that are in their first excited quantum state ("occupied once" – with all others in their ground state), a quasi-classical *field* state has in QFT to be described as a coherent superposition of many *different* excitations $\psi_i^{(n)}(q_i,t)$ (different "photon numbers" $n$) for each spatial eigenmode $i$. In contrast to the *free* wave packet shown in Fig. 1, these "coherent oscillator states" (time-dependent Gaussians, now functions of the field amplitude) preserve their shape and width *exactly*, while their centers follow classical trajectories $q_i(t)$. Therefore, they imitate oscillating classical fields much better than wave packets in space may imitate particles.

One and the same quantum concept of field functionals $\Psi$ may thus consistently represent "complementary" classical concepts such as "particle" numbers and field amplitudes – albeit again mutually restricted by a Fourier theorem. For this reason, the *Boltzmann distribu-* *tion* e$^{-E/kT}$ of their energy eigenstates may describe the Planck spectrum with its particle and wave limits for short and long wavelengths, respectively. Field functionals can also describe all specific phenomena of quantum optics, such as "photon bunching".

## 4. Entanglement and Quantum Measurements

Before trying to study *interacting* quantum fields (Sect. 5), early quantum physicists successfully investigated the quantum mechanics of non-relativistic "many-particle" systems, such as multi-electron atoms, molecules and solid bodies. These systems could often *approximately* be described by means of different (orthogonal) single-particle wave functions for each electron, while the atomic nuclei seemed to possess fixed or slowly moving positions, similar to classical objects. For example, this picture explained the periodic system of the chemical elements. On closer inspection it turned out, however, – at first for the ground and excited states of atoms and small molecules – that *all N* particles forming such objects, including the nuclei, have indeed to be described by one common wave function in their *3N*-dimensional configuration space. This cannot normally be a product or determinant of single-particle wave functions – a consequence that was later called "entanglement". It must similarly apply to different wave modes $q_i$ of interacting fields in QFT. Given any two systems, the set of all their separating (non-entangled) quantum states can only have measure zero. When David Bohm began to study consequences of this fundamental property for his theory of 1952, he referred to it as "quantum wholeness", since entanglement means that quantum theory can *only* be consistently understood as quantum cosmology (see Sect. 6). Historically, the essential role of this generic entanglement was often belittled as a mere statistical correlation between subsystems, while this misinterpretation was then incorrectly used in turn as an argument against an ontic interpretation of the wave function. The presently very popular toy model of entangled qubits, interrupted by classically described actions of Alice and Bob, is but an inconsistent caricature of quantum mechanics.

Every physics student is using the entanglement between an electron and a proton in the hydrogen atom when writing the wave function as a product of functions for center-of-mass and relative coordinates. This would not make sense for interacting fields, which are always separately defined. While the wave function for the relative coordinates then defines the size and structure of the hydrogen atom, the center of mass may be represented by a free spatial wave packet as in Fig. 1. The simplest nontrivial case of entanglement, the Helium atom, was first successfully studied in great numerical detail by Hylleraas, using variational methods, in a series of papers starting in 1929. Already Arnold Sommerfeld noticed in his *Wellenmechanischer Ergänzungsband* that "Heisenberg's method", which used only the anti-symmetrization of product wave functions by means of "exchange terms", is insufficient to correctly describe multi-particle systems. (Anti-) symmetrisation is often confused with entanglement, since it formally describes entanglement between physical variables and meaningless particle numbers, that merely eliminates any concept of distinguishability.† It is therefore not required any more in the occupation number representation of QFT (see Sect. 5). For long-range interactions, entanglement may be small in low-energy states, since it corresponds to "virtual excitations" of single-particle states (which are often misinterpreted as "fluctuations" rather than static entanglement).

An important consequence of entanglement is that subsystem Hamiltonians may be not (or not uniquely) defined – thus excluding *local* unitarity and a uniquely defined Heisenberg or interaction picture for open systems. *Closed* non-relativistic *N*-particle systems, on the other hand, have to be described by *one* common wave function in their complete configuration space. Their center-of-mass motion may then factorize from the rest, thus leading to a *spatial* wave function for it (identical to that for mass points or "quantum particles"), while the internal energy quantum numbers of such systems are given by the numbers of nodes (now forming hypersurfaces) in the remaining *3(N-1)*-dimensional configuration space. For non-inertial motion, this separation holds only approximately.[5] Although approximately closed systems represent an exception, they were the main objects being studied for simplicity while quantum theory was being developed. Open system quantum mechanics was invented much later – mostly in combination with statistical physics. Time-dependent Hamiltonians are a relic of classical physics, too, as they would require time-dependent separating states as a source of the time-dependent forces (thus neglecting entanglement). When unitary dynamics was consistently applied to a *global* system in order to derive subsystem quantum dynamics, it led to local non-unitarity, and in particular to the concept of decoherence.

However, how can the space of all possible classical configurations, which would even possess varying dimensions, in this way replace three-dimensional space as the fundamental arena for the dynamics of physical states? If our Universe consisted of $N$ particles (and nothing else), its configuration space would possess $3N$ dimensions – with $N$ being at least of order $10^{80}$. For early quantum physicists – including Schrödinger, of course – such a wave function was inconceivable, although the concept of a space of *possible configurations* fits excellently with Born's probabilities for classical properties. Entanglement can then conveniently be understood as describing statistical correlations between measured variables. But only between *measured* variables! Since macroscopic variables are "permanently measured" by their environment (see below for decoherence), their entanglement almost always appears as no more than a statistical correlation. This explains why we are used to interpret the space on which the wave function is defined as a "configuration" space, even though entanglement is responsible, for example, for the precise energy spectrum and other *individual* properties of bound microscopic systems – regardless of any statistical interpretation. This conceptual difference is often "overlooked" for convenience in order to keep up the illusion of an epistemic interpretation of the wave function (where probabilities would reflect incomplete information about "hidden" variables). Even after individual scattering events one often needs entangled scattering amplitudes with well-defined phase relations between all fragments, while mere scattering *probabilities* would be insufficient to describe the situation. But only after Einstein, Podolsky and Rosen (EPR) had shown in 1935 that the entanglement between two particles at a distance may have directly observable consequences, did Schrödinger regard this property as the greatest challenge to his theory – although he kept calling it a "statistical correlation". EPR had indeed erroneously concluded from their analysis that quantum mechanics cannot represent a complete description of Nature, so that hidden variables had to be expected.

---

† *Separate* (anti-)symmetrization of spin and orbit parts, however, may define *physical* entanglement between particles. Genuine entanglement in many-electron atoms would mean, for example, that one has to take into account "configuration mixing" in jj-coupling as a correction to the independent-particle (Hartree-Fock or mean field) approximation. Therefore, the statement "particle at position $x_1$" (in contrast to "particle number 1") "has spin-up" – as in a Bell type experiment – is *physically* meaningful.

While many physicists speculated that such hypothetical hidden variables could possibly never be observed (even though they might exist), it came as a surprise to them when John Bell showed in 1964 that *any* kind of hidden local reality (no matter whether it consists of particles, fields or other local things with local interactions only – observable or not) would be in conflict with certain observable consequences of entangled wave functions. This conclusion eliminated the most important argument for an epistemic interpretation of the wave function (namely, the possibility of a statistical explanation of its suspicious non-locality). In order to prove his theorem, Bell used arbitrary local variables $\lambda$ (just a name for something not yet known) for an indirect proof, but most physicists had by then become so much accustomed to Bohr's rejection of microscopic reality that they immediately accused Bell for having used a "long refuted assumption". However, Bohr had been led to his radical denial of microscopic reality only because of his insistence on classical concepts (such as particles), and since Hei-senberg's original justification of his uncertainty principle by means of unavoidable *perturbations* of conjugate variables during measurements had turned out to be insufficient. Bohr's philosophical position should therefore not be used as an argument against an ontic interpretation *of the nonlocal wave function*. No logic can exclude the possibility that apparent points in configuration space (classical "states") are in reality narrow wave packets.

Crucial direct tests of Bell's consequence of quantum nonlocality had in practice to be restricted to two- or few-particle systems, which can be isolated from everything else until they are measured. While their entanglement, which is a direct consequence of the superposition principle, has thereby always been confirmed, physicists are still debating whether this fact excludes locality or any kind of "microscopic reality" (which may just mean the existence of a complete and consistent formal description). None of these two sides feels particularly bothered by Bell's theorem. The two camps usually prefer the Schrödinger picture (in terms of wave functions) or the Heisenberg picture (in terms of observables), respectively, and this seems to be the origin of many misunderstandings between them. In the absence of any local *states*, dynamical locality ("relativistic causality") may even appear difficult to define – but see the discussion in the third paragraph from the end of Sect. 5: QFT *is* dynamically local even in the relativistic sense.

If one does assume the superposition principle to apply universally, one is forced to accept one entangled wave function for the whole universe. Heisenberg and Bohr assumed instead that the wave function is no more than a calculational tool which "loses its meaning" after the final measurement that concludes an experiment. This "end of the experiment" (related to the "Heisenberg cut") remains vaguely defined and *ad hoc*. Its traditional application (namely, too early in the chain of interactions that leads to an observation) had indeed delayed the discovery of decoherence, which will be discussed below, for several decades. When I first suggested its importance, I was regularly told by colleages that quantum theory does not apply to the environment. A universal wave function that always evolves according to the Schrödinger equation, however, leads to an entirely novel world view that, in spite of being entirely consistent, still appears inacceptable to many physicists.

For example, if one measures a microscopic object that is initially in a superposition of two or more different values of the measured variable, this gives rise to an entangled state for the microscopic system and the apparatus – the latter including Schrödinger's infamous cat if correspondingly prepared. (All unitary interactions discussed here and below may be assumed to be of a form like $(\sum_n c_n \psi_n)\Phi_0 \rightarrow \sum_n c_n \psi_n \Phi_n$, that is, transforming local superpositions into en-

tanglement – thus in the "ideal" case without changing or "disturbing" the measured states $\psi_n$.) Since superpositions of different pointer positions have never been observed, one traditionally (but perhaps unnecessarily) postulates according to von Neumann that Schrödinger's dynamics has to be complemented by a stochastic "collapse of the wave function" into one of its components, or into a product of narrow wave packets for all macroscopic variables. However, in a unitary description there is no distinction between measurements and general interactions. Heisenberg's "observables" are then readily defined (up to a scale) by the interaction Hamiltonian between system and apparatus rather than forming an independent ingredient of the theory. Since this interaction characterizes a measurement device regardless of the time of its application, it appears quite unreasonable to endow it with the actual dynamics *of the object* (as done in the Heisenberg picture). In the Copenhagen interpretation, one would pragmatically jump from a description in terms of wave functions to one in classical terms, and back to a new wave function in order to describe a subsequent experiment. This unsatisfactory situation is known as the *quantum measurement problem*.

If one is ready, instead, to accept a universal Schrödinger equation to describe reality, one has to understand what an entangled wave function for the microscopic system plus an apparatus would mean. To do so, one has to include the observer into the quantum description.[6] When he reads off the measurement result, he becomes himself part of the entanglement. According to the unitary dynamics, he would thereafter simultaneously exist in different states of awareness (different states of mind) – similar to the fate of Schrödinger's cat. Hugh Everett first dared to point out in 1957 that this consequence is not in conflict with our subjective observation of *one* individual outcome, since each arising "component state" (or "version") of the observer can register and remember (hence be aware of) only that outcome which is realized in his corresponding "relative state" of the world. This partial state would also contain only compatible versions of all the observer's "friends" – thus defining *objectivized* outcomes. As there are many such correlated component states (with their many minds) in one global superposition, though, the question which of them contains the physicist who prepared the experiment has no unique answer; according to the unitary dynamics they all do.

Why can these components be regarded as separate "worlds" with separate observer states? The answer is that they are dynamically "autonomous" after an irreversible measurement in spite of their common origin; each of them describes a quasi-classical world for its macroscopic variables (see the discussion of decoherence below). In contrast to identical twins, who also have one common causal origin, different versions of the "same" observer in autonomous branches cannot even communicate according to the unitary dynamics, and thus can conclude each other's existence only by means of an extrapolation of the dynamical laws that they happen to know. This is certainly an unconventional, but at least a consistent picture, since it is a straightforward consequence of the Schrödinger equation; it only requires the recognition and identification of some semi-autonomous physical "systems", whose definition is consistent with a global nonlocal wave function under local interactions, with subjective observers. Attempts to avoid this conclusion are all motivated by traditional prejudice, and would lead back to an unsolved measurement problem.

Many physicists still prefer to believe, instead, that some conceptual or dynamical border line between micro- and macrophysics must exist – even though it could never be located in an experiment. Otherwise it should be possible (so it seemed) to observe individual consequences of entanglement between microscopic systems and their macroscopic measurement instruments – similar to the energy or other properties of Hylleraas's entangled Helium atom or of small molecules. However, this bipartite entanglement is not yet complete. Macroscopic systems must inevitably and extremely fast interact with their natural "environment", whereby the entanglement that had resulted from the measurement proper would uncontrollably and irreversibly spread into much of the "rest of the universe". This happens even before an observer possibly enters the scene. In this way, one may understand how a superposition that extends over different macroscopic pointer positions, for example, would, from the point of view of a potential local observer, inevitably be transformed into an *apparent ensemble* of narrow wave packets that mimic classical states (points in configuration space) as potential outcomes. Although still forming one superposition, these partial waves, which may include different states of all observers, have no chance to meet again in high-dimensional configuration space in order to have local coherent consequences. In this sense (only), they can now be *regarded as forming an ensemble* of different "worlds".

This unavoidable entanglement with the environment (whose onset defines the true border line between micro- and macrophysics) is called decoherence,[7] as predominantly phase relations defining certain quantum mechanical superpositions become locally unavailable: they are irreversibly "dislocalized".[‡] As Erich Joos and I once formulated it, the superposition

---

[‡] A mere phase *randomization* ("dephasing") could neither occur under unitary dynamics, nor would it solve the issue, as each individual member of an ensemble of superpositions with different phases would remain a superposition (though possibly with *unknown* phase). Similarly, local phases that are assumed to fluctuate rapidly in time for some reason are in a definite superposition at any instant. Nonetheless, phase averaging forms the most popular misunderstanding of decoherence, which describes entanglement with the environment in the *individual* case (no averaging). These different concepts are easily confused, in particular, if the environment is described

still exists, but it "is not there" (somewhere) any more. Decoherence is in general a very drastic consequence of quantum dynamics, which requires (and allows) precise numerical calculations only for some mesoscopic systems, such as chiral molecules.[8] For example, some of the latter are found in chiral states on Earth, but in parity eigenstates in interstellar space, where interaction with the environment is weaker. In contrast, all objects that can be "seen" under normal conditions must permanently scatter light, which is thus entangled with the state of the object. If two positions can be distinguished by "just looking", the quantum states of their scattered light must be orthogonal, and thus decohere the object. The position basis is here "preferred" by the locality of interactions.

The time asymmetry of the decoherence process (in *causing* uncontrollable entanglement) can be explained by means of a low entropy initial condition for the global wave function.[9] However, without the other non-trivial consequence of global unitarity, namely the concept of splitting observers ("many minds"), decoherence would not be able to explain the observation of *individual* measurement outcomes. Decoherence thus represents only half of the measurement story, while Everett without decoherence would be ambiguously defined. Decoherence has therefore occasionally been questioned even to contribute the quantum measurement problem, but the subsequent splitting of observer states amounts for the latter to what Pauli once called the "creation of measurement results outside the laws of Nature"; it is now described as a dynamical consequence of global unitary dynamics *on the observer himself*. Instead of properly taking into account the environment and the role of the observer in a consistent quantum setting, that is, in a deeply entangled world, Pauli, Heisenberg and their disciples had to refer to an extra-physical observer or his "information" (a concept of questionable popularity) as a *deus ex machina*.

---

as a "thermal bath". However, if this initial thermal "mixture" had been caused by earlier *quantum* interactions with the environment (which is its most realistic origin in a quantum world), the thus pre-existing entanglement would simply be dynamically extended to the "dephased" variables, where it would then also lead to their genuine decoherence (a dislocation of their *individual* relative phases). Using the reduced density matrix formalism would tacitly replace nonlocal entanglement by local ensembles: entanglement is insufficiently defined for "mixed states". It is remarkable that many important physicists are still missing the essential point of decoherence as a consequence of the fundamental nonlocality of (pure) quantum *states*. Nonlocal phase relations may even define observable individual properties in microscopic systems (such as the total spin of two particles at different positions), but nonetheless contribute to the decoherence of their subsystems. – Historically, the term "decoherence" was first invented in the context of "decoherent histories" in about 1985, where it was *postulated* in order to justify "consistent histories" within a conventional probability interpretation, whereas my arguments of 1970 were derived from the assumption of universal unitarity in an attempt to *resolve* the measurement problem – not in order to tolerate it. Ironically, it is precisely decoherence as a consequence of universal unitarity that had led to the traditional prejudice that quantum theory does *not* apply to the macroscopic world.

The experimental confirmation of decoherence as a smooth (though very fast) dynamical process[4] clearly demonstrates that the concept of entanglement does apply beyond microscopic systems. This unification of two apparently separate realms of physics (quantum and classical) was certainly the most important achievement of decoherence theory. While this process must remain uncontrollable in order to be irreversible (thus giving rise to the "real" rather than "virtual" appearance of its outcome), it has many obvious and important consequences – including apparent quantum jumps and the classical appearance of the world (as consisting of particles *and* fields). So it explains why we seem to observe individual atoms as apparent particles in a Paul trap, or tracks in a Wilson chamber as apparent particle trajectories (both are correctly described in terms of narrow wave packets), and why one finds bound microscopic systems preferentially in their energy eigenstates.[7,10] It also allows us to understand the mysterious concept of "complementarity" simply by the different entanglement of microscopic objects with the environment, caused by different measurement instruments. This choice of "complementary measurement devices" is not available any more for systems, such as macroscopic ones, that are already strongly entangled with their unavoidable environment – without being measured by a physicist. The basis "preferred" by this unavoidable environment then defines a quasi-classical *configuration* space for such systems, which include even major parts of the thus partially classical observers (such as their neural system).

Decoherence also explains Dirac's famous remark that "a photon can only interfere with itself". This statement appears surprising since all "photons" are excitations of one and the same quantum field (cf. Sects. 3 and 5), and are thus known to be indistinguishable. However, if two equal atoms at positions $A$ and $B$, say, are initially in the same excited state, they may independently decay by emitting an outgoing radiation mode centered at their respective position. Since these two wave modes are then correlated with different (orthogonal) final states of their total source, they are decohered from one another, and so appear to represent "different photons". If the excited atoms were replaced by lasers as radiation sources, these coherent states would *not* change when emitting a photon, and even the superposition of wave packets emitted by different lasers may interfere where and whenever they overlap; they represent "one and the *same* photon". The relative phase between these two components may initially not be known, but evidently it *exists* and can be observed (cf. Footnote ‡). This superposition of two photon components has to be distinguished from a "two-photon" state (double-decay), which has *two* nodes in its field functional, may be described by a symmetrized product wave function, and will cause *two* clicks in the detector.

While *virtual* decoherence had always been known in the form of microscopic (reversible and often observable) entanglement, the unavoidable and irreversible consequence of the environment on macroscopic systems was overlooked for five decades, mainly because quantum mechanics was traditionally assumed *not* to apply beyond microscopic systems. Surprisingly, the apparently reversible classical mechanics requires in unitary description the permanent (though mostly thermodynamically negligible) action of irreversible decoherence.

In order to illustrate the enormous number of new parallel "worlds" that are permanently created by means of decoherence (or would otherwise be permanently annihilated by a collapse mechanism), let me consider the example of a two-slit experiment. Measuring which slit the "particle" passes would double the number of worlds, but registration of the particle on the second screen causes a multiplication of worlds by a large factor that depends on the remaining coherence lengths for the positions of the decohered spots. (Everett "worlds" need not be *exactly* orthogonal, and thus cannot simply be counted; they may even form an overcomplete set.) This definition of branch worlds by their irreversible separation in configuration space means also that quantum computers do *not* simultaneously calculate in parallel worlds (as sometimes claimed) if they are to produce a coherent result that may then be used in "our" world, for example; "real" (rather than virtual) branches can by definition not lead to observable (local) superpositions anymore.

Most "particles" in the two-slit experiment do not even pass the slits, but may instead be absorbed on the first screen. This would correspond to a position measurement, too – regardless of whether its information is ever extracted. In order to cause decoherence, this "information" may even be thermalized (erased in the usual sense). In contrast, a "quantum eraser" requires a local superposition to be *restored*, that is, re-localized, rather than information to be destroyed, as its name may suggest. Similar irreversible entanglement forms in all occasional interactions between different quantum systems. For $M$ such "measurement-like events" in the past history of the universe with, on average, $N$ different outcomes, one would obtain the huge number of $N^M$ now existing branches. Nonetheless, the global configuration space remains almost empty because of its huge dimension; the myriads of branching wave packets that have ever been created by *real* decoherence describe autonomous "worlds" for all reasonable times to come. Nobody can calculate such a global wave function, of course, but under appropriate (far from equilibrium) initial conditions for the universe, its unitary dynamics can be used consistently to justify (1) quasi-classical properties and behavior for all degrees of freedom that are "robust" under decoherence, (2) statistical methods (retarded proba-bilistic master equations) for most others,[9] and (3) individual wave functions for appropriately prepared microscopic systems. In the case of controllable non-local entanglement, this latter kind of preparation can even be applied at a distance – a phenomenon known as "quantum steering". These three dynamical applications are then also sufficient to describe measurement devices to begin with. No phenomenological concepts (such as particles, events, pointer positions, or even Alice and Bob) are required on a *fundamental* level.

The observation of radioactive decay represents another measurement of a continuous variable (namely, the decay time). Its precision cannot be better than the remaining coherence time (which is usually very much smaller than the half-life, and thus gives rise to apparent quantum jumps). This coherence time depends on the efficiency of the interaction of the decay fragments with their environment, and it would be further reduced by permanent registration of the (non-) decay. If an excited state decays only by emission of weakly interacting photons, however, decoherence may be relatively slow. In a cavity, one may then even observe a superposition of different decay times, thus definitely excluding genuine quantum jumps ("events") in this case. There is no reason to believe that this would be different if the photon had travelled astronomical distances before such a coherent state vector revival occurs.

Many leading physicists who are not happy any more with the Copenhagen interpretation nonetheless prefer to speculate about some novel kind of dynamics (an as yet unknown collapse mechanism) that would avoid the consequence of Many Worlds. This is at present no more than prejudice or wishful thinking, but it could in principle also solve the measurement problem in terms of an ontic (in this case partially localized) universal wave function without requiring Everett's splitting observers. One should keep in mind, though, that all as yet *observed* apparent deviations from unitarity, such as quantum jumps or measurements, can be well described (and have in several cases been confirmed experimentally) as smooth decoherence processes in accordance with a global Schrödinger equation. Therefore, if a genuine collapse mechanism did exist after all, it would presumably have to be *triggered* by decoherence, but it could then hardly have any observable consequences on its own.

For example, if one of two spatially separated but entangled microscopic systems (such as those forming a "Bell state") was measured, their total state would according to a unitary description become entangled with the apparatus, too, and thus also with the latter's environment. While this process leads to the formation of two dynamically autonomous branches, an observer at the location of the second system, say, becomes part of this entanglement (and therefore "splits") only when he receives a signal about the result. Before this happens, his state factors out, and he may be said not yet to *know* the result. Since the local mixed state resulting from decoherence represents an apparent ensemble, this final step of an observation *appears* as a mere increase of information about an already existing property. If he also measured the second system (that at his own location in this example), the state of his memory must thereafter depend on the outcomes of both measurements, that is, it must have split twice unless there was an exact correlation between the results. Since the order of these two measurements does not matter, in general, this description includes delayed choice experiments. In contrast, a genuine collapse caused by the measurement would have to *affect* distant objects instantaneously (whatever that means relativistically) in order to avoid other weird consequences. This *would* then define the "spooky" part of the story.

An *apparent ensemble* of quasi-classical "worlds" is not exactly, but for all practical purposes ("FAPP") sufficiently, defined by the autonomous branches of the wave function that arise from decoherence: a measurement cannot be undone in practice as soon as the global superposition cannot be re-localized any more. Reasonable observer states can then only evolve separately within the different branches. Neither can we as yet precisely define conscious observer systems in physical terms, nor would this definition completely explain Born's rule: observers in many branches would, in series of measurements, even observe frequencies of outcomes that are *not* in accord with Born's rule. What we still need, therefore, is a probabilistic characterization of the quasi-classical world in which "we" happen to live.

In all interpretations of quantum mechanics, Born's rule had to be *postulated* (in addition to the unitary dynamics) on empirical grounds in some form. In the Everett interpretation, the situation is now partly solved by the unitary mechanism of decoherence, as the members of an *effective ensemble* of potential physical "outcomes" are sufficiently defined. In contrast to the splitting of *subjective* observers, this branching into autonomous "worlds" is not a new fundamental concept; it is a consequence of global unitary dynamics. According to their definition by robustness against further decoherence, there are no autonomous branches that contain Schrödinger cats, sugar molecules in parity eigenstates, or other "macroscopic" superpositions. So their probability is zero. All we still have to postulate for the remaining branches are probability *weights* for subjective observers (us) to find ourselves existing in them. For these weights to be dynamically consistent, the squared norm (the *formal* measure of size of the branches) is the only reasonable candidate, since it is additive and conserved under unitary dynamics, and thus not affected by any subsequent finer branching. (For example, further branching occurs during subsequent physical information processing, such as photon "meas-urements" on the retina, or by measurement-type events somewhere else in the universe.) Therefore, these weights give rise to individual probabilities for apparent collapse *events*, and thus to the concept of "consistent histories". Everett regarded this dynamical argument, which is similar to the choice of phase space volume in classical mechanics, as *proof* of Born's probabilities.[11] However, only *after explicitly postulating* them, does the density matrix (called a "mixed state") become justified as a conceptual tool, as it allows us to calculate all "expectation values" according to these probabilities.

By consistently using this global unitary description, all those much discussed "absurdities" of quantum theory can be explained. It is in fact precisely how they were all predicted – except that the chain of unitary interactions is usually cut off *ad hoc* by a collapse at the last relevant measurement in an experiment, where the corresponding decoherence defines a consistent position for the hypothetical Heisenberg cut. Therefore, all those "weird" quantum phenomena observed during the last 80 years can only have surprised those who had never seriously considered the possibility of a universal validity of unitarity (including most of the founders of quantum theory). Absurdities, such as "interaction-free measurements", arise instead if one assumes the quasi-classical phenomena (such as events) rather than the wave function as describing "reality". If the wave function itself represents reality, however, any "post-selected" component would not describe the previously documented past any more, as it should be the case if this post-selection was no more than a "normal" increase of information.

So-called quantum teleportation is another example where one can easily show, using unitary dynamics, that nothing is ever "teleported" that, or whose deterministic predecessor, was not prepared *in advance* at its intended position in one or more components of an entangled initial wave function.[10] This confirms again that entangled wave functions cannot merely represent a bookkeeping device – even though a local observer *may pretend* that an objective global collapse into a non-predictable branch had already occurred (or that the outcome had been *created* in some other kind of "event") as a consequence of the first irreversible decoherence process after a measurement. It is precisely this possibility that justifies the usual pragmatic approach to quantum mechanics (including Bohr's Copenhagen interpretation or von Neumann's collapse during a measurement). However, if one assumed only local properties, such as quasi-classical measurement outcomes, to describe reality, one would indeed have to believe in teleportation and other kinds of spooky action at a distance. According to the Everett interpretation, the usual restriction of "our" quantum world to a tiny and permanently further collapsing *effective* wave function therefore represents no more than a pragmatic conven-tion that reflects the observer's changing physical situation rather than a collapse process. Such a "collapse by convention" could even be assumed to apply instantaneously (superluminally), but it should be evident that a mere convention cannot be used for sending signals.

If the global state does indeed always obey unitary dynamics, the observed quantum indeterminism can clearly *not* represent any objective dynamical law. In the Everett interpretation, it is in principle a "subjective" phenomenon that reflects the branching histories of all observers into many different versions (with their "many minds"). This may *explain* Heisenberg's original interpretation of quantum measurements as requiring "human" observers. This passive indeterminism is nonetheless essential for the *observed* dynamics of the quantum world (some observers' "relative state"). All measurement outcomes are objectivized by the correlation between those versions of *different* observers (including Wigner's friend or Schrödinger's cat) who exist in the same Everett branch, and thus can communicate. For all practical purposes, their entanglement with the apparatus after reading it, and with the environment, also justifies Bohr's interpretation of measurements (unlike Heisenberg's) in terms of classical outcomes that would be irreversibly and *objectively created* (in apparent events) by the macroscopic apparatus. This macroscopic entanglement (in addition to decoherence) explains the traditional concept of a causal "classical reality": only a documented phenomenon is a phenomenon (see also Footnote ** in Sect. 5). Only if one misinterpreted the resulting global superposition as a statistical ensemble with respect to an *external* observer, would an observation of the outcome appear as a mere increase of the latter's information.

### 5. Quantum Field Theory

We have seen that quantum mechanics in terms of a universal wave function admits a consistent (even though novel kind of) description of Nature, but this does not yet bring the strange story of particles and waves to an end. Instead of *spatial* waves (fields) we were led to wave functions on a high-dimensional "configuration space" (a name that is justified only by its appearance as a space of *potential* classical states because of decoherence). For a universe consisting of $N$ particles, this configuration space would possess $3N$ dimensions, but we may conclude from the arguments presented in Sect. 3 that for QED (quantum electrodynamics) it must be supplemented by the infinite-dimensional configuration space of the Maxwell fields (or their vector potentials in the canonical formalism). A product of wave functions for the amplitudes of all field modes in a cavity or in free space turned out to be sufficient to explain

Planck's quanta by the number of nodes of these oscillator wave functions. The spontaneous occurrence of photons as apparent particles (in the form of clicking counters, for example) is then merely a consequence of the fast decoherence caused by the macroscopic detector.

However, we know from the quantum theory of relativistic electrons that they, too, have to be described by a *quantized field* (that is, by a field functional) – a consequence that must remain true in the non-relativistic limit. There are then no particles even *before* quantization. The relativistic generalization of a one-electron wave function is called the *Dirac field*, since it is usually studied as a function on spacetime. Dirac proposed it successfully for the hydrogen atom at a time when Schrödinger's wave function was still believed to define a spatial field for each electron, but the Dirac field can *not* be generalized to an *N*-electron field on a *4N*-dimensional "configuration spacetime", although this has occasionally been proposed. There is only one time parameter describing the dynamics for the *total* state. In the Schrödinger picture of QED, the Dirac field is used to define, by its configuration space and that of the Maxwell field, the space on which the corresponding time-dependent wave functionals live. According to the rules of canonical quantization, these wave functionals have to obey a generalized Schrödinger equation again (also called the Tomonaga equation).[12]

Spin and other internal degrees of freedom thereby become part of the "classical" (not-yet-quantized) spatial fields. While such position-dependent spinors then form a kind of "classical entanglement" between position and spin of one object, they do not imply any Bell type non-locality. For example, the dynamical separation of spatial wave packets corresponding to different neutrino masses that contribute to their initial flavor while the neutrinos travel astronomical distances, has also been called "decoherence". Since it may be regarded as the longitudinal version of a Stern-Gerlach (pre-) measurement of the spin by the position variable, this neutrino decoherence should also be called "virtual", even though the neutrino wave packets can hardly ever be recombined in practice in order to restore the initial flavor. Real decoherence is usually defined as the uncontrollably arising entanglement with an unbounded number of unavoidable environmental degrees of freedom – cf. also footnotes in Sect. 4 – but this is a matter of definition. The recent detection of *coherent neutrino scattering*[13] demonstrates indeed that individual neutrinos have to be described by field modes (just as photons), while their apparent particle properties are again caused by genuine decoherence (an apparent collapse into a wave packet) in their detectors *and* (for neutrinos) also during their creation process (which gives rise to incoherent ensembles of radial wave packets or emission times).

Nonrelativistically, the formalism of QFT avoids an *N*-dependence of the required configuration spaces for different numbers *N* of "particles". Quite generally, it allows for a concept of "particle creation", such as by raising the number of nodes of the field functional (cf. Sect. 3). Relativistic covariance cannot and need not be manifest in the canonical formalism. For example, the canonical quantization of the Maxwell field leads consistently to a wave functional *Ψ{A(x);t}*, with a vector field *A* defined at all space-points *x* on an arbitrary simultaneity *t*. Since Schrödinger had originally discovered his one-electron wave function by the same canonical quantization procedure (applied to a single mass point), the quantization of the Dirac field is for this purely historical reason also called a "second quantization". As explained in Sect. 4, though, the particle concept, and with it the first quantization, are no more than historical artifacts.[14]

Freeman Dyson's "equivalence" of using relativistic field functionals (Tomonaga) or time-dependent field operators (Feynman)[15] is essentially based on the limited equivalence between the Schrödinger and the Heisenberg picture. The latter would hardly be able even in principle to describe the hefty, steadily growing entanglement described by a time-dependent global wave functional. It is therefore mostly restricted to the quantization of *free* fields (coupled oscillators, which can easily be quantized algebraically). Since relativity is incompatible with absolute simultaneities, the relativistic generalization of the Schrödinger equation can only be given by the Tomonaga equation with its "many-fingered" concept of time (arbitrary simultaneities). Apparent particle lines in Feynman diagrams, on the other hand, are merely shorthand for free field modes (such as plane waves, with "particle momenta" representing their wave numbers).[3] These diagrams are used as intuitive tools to construct terms of a perturbation series described as integrals over products of such *field modes* and other factors – mainly for calculating scattering amplitudes. In this picture, closed lines ("virtual particles") may represent local entanglement between quantum fields. Since high-energy physics is mostly restricted to scattering experiments, unitarity is in many textbooks quite insufficiently defined by the conservation of probability – thus neglecting its essential consequence for the quantum phases, which are needed to determine nonlocal superpositions that must arise in such a scattering process. Neglecting them would *presume* a probabilistic collapse.

The Hamiltonian form of the Dirac field equation is unusual because of its linearization in terms of particle momentum: the classical canonical momenta are not given by time derivatives of the position variables (velocities) any more. Nonetheless, the two occupation numbers 0 and 1 resulting from the *assumption* of anti-commuting field operators[§] are again interpreted as "particle" numbers because of their consequences in the quasi-classical world. Field modes "occupied" once in this sense and their superpositions define again "single-particle wave functions". In contrast to the case of photons, however, one never observes superpositions (wave functionals) of *different* electron numbers. This has traditionally been regarded as a fundamental restriction of the superposition principle (an axiomatic "charge superselection rule"), but it may be understood as another consequence of decoherence: for charged particles, their Coulomb field assumes the role of an environment.[16] While there are proposals to explain discrete quantities, such as charge, as "winding numbers" for some field variables, this would not exclude their quantum superpositions.

In QFT, the formulation that one particle is in a quantum state described by the spatial wave function $\psi_1$, and a second one in $\psi_2$, is thus replaced by the statement that two *field modes*, $\psi_1$ and $\psi_2$, are both in their first excited quantum state ("occupied once"). A permutation of the two modes does not change this statement, which is based on a logical "and". So there is only *one* state to be counted statistically. This eliminates Gibbs' paradox in a very natural way. (Schrödinger seems to have used a similar argument in favor of waves instead of particles even before he explicitly formulated his wave equation.[17])

It would similarly be inappropriate to claim that *wave functions* can be directly observed in Bose-Einstein condensates (as is often done). What one does observe in this case are again the (now many times "occupied") three-dimensional boson *field modes*. This includes massive bosons, which are traditionally regarded as particles. Instead of the *free* field modes used for photons, however, interacting bosons are then more appropriately described in terms of self-consistent field modes, similar to the self-consistent Hartree-Fock single-fermion wave functions. Both methods neglect any "particle" entanglement, and can therefore at most represent approximations for the lowest states. They lead to what is regarded as an effective non-linear "single-particle wave equation" – for bosons called the Gross-Pitaevskii equation.[**] In spite of this effective non-linearity, the quantum states proper are, of course, still described by the linear Schrödinger equation – relativistically always in the sense of Tomonaga.[12]

---

[§] Let me emphasize, though, that the origin of the Pauli principle, which is valid for fermions, does not seem to be entirely understood yet. While the individual components of the Dirac spinor also obey the Klein-Gordon equation, the latter's quantization as a field of coupled oscillators would again require *all* oscillator quantum numbers $n = 0,1,2,...$. Anti-commuting field operators, which lead to anti-symmetric multi-particle wave functions under permutations, were postulated quite *ad hoc* by Jordan and Wigner, and initially appeared artificial even to Dirac. Interpreted rigorously, their underlying configuration space (defining a Hilbert space basis again) would consist of a spatial continuum of bits ("empty" or "occupied") rather than a continuum of coupled oscillators. The $n$-th excited state of this bit continuum (that is, $n$ occupied positions) represents $n$ *identical* point-like "objects". Because of the dynamical coupling between bit-neighbors, these objects can move, but only *after* their quantization, which leads to entangled superpositions of different occupied space points, may they give rise to propagating waves. In order to be compatible with this bit continuum, the coefficients of these superpositions ("multi-fermion wave functions") must vanish whenever two of their arguments coincide. This can quite generally be achieved by assuming them to be antisymmetric under permutations of any two arguments. No field algebra is explicitly required for this argument (although it could then be consistently defined). In this picture, single-fermion wave functions would represent genuine quantum states (quantum superpositions) rather than wave modes as for bosons. In contrast, coupled oscillators defining a free boson field propagate as spatial waves, and thus obey a *classical* superposition principle (in space rather than in their configuration space) in addition to the quantum superposition principle that is realized for them by the field functionals. This difference would be particularly dramatic in Bohm's theory, where one often meets disagreement on whether its trajectories have to include photons as particles or as a time-dependent vector potential (a classical field). However, these pre-quantization concepts need not possess any physical meaning by themselves. Moreover, such a fundamental distinction between bosons and fermions may be problematic for *composite* "particles" (dressed fields).

As already mentioned in Sect. 3, photon number eigenfunctions $\psi^{(n)}(q)$ in the configuration space of wave amplitudes $q$ – to be distinguished from their three-dimensional field modes ("single-photon wave functions", which are fixed modes in a cavity in this case) – have recently been observed by means of their Wigner functions, and thus confirmed to exist, for various values of the "particle number" $n$.[4] For pure states, Wigner functions are defined as partial Fourier transforms of the dyadic products $\psi^{(n)}(q)\psi^{(n)*}(q')$, and thus equivalent to the wave functions $\psi^{(n)}(q)$ themselves (except for a total phase). The variable $q$ is here the amplitude of the given field mode rather than some spatial position as in single-particle quantum mechanics. The two-dimensional Wigner functions on their apparent phase space $q,p$ were made visible in this experiment, and so allow one to clearly recognize the $n$ nodes of the wave functions $\psi^{(n)}(q)$ (forming circles in phase space). Creation and annihilation operators are defined to change the number of these nodes. Since these operators occur dynamically only in the Schrödinger equation, they describe *smooth* physical processes (time-dependent wave functionals), while creation "events" are either meant just conceptually, or would require a fast decoherence process. The *physical* nature of field functionals is also confirmed by their ability to participate in the general entanglement of interacting systems and, in this way, contribute to the observable decoherence of the states of radiation sources without having to affect any "real" absorbing matter (thus also excluding any action at a distance in this case).

---

[**] At higher temperatures, "many-particle" systems (that is, multiple quantum field exitations) may behave approximately like a gas of classical particles undergoing stochastic collisions because of the mutual decoherence of the field modes into apparent ensembles of narrow spatial wave packets.[29] This consequence perfectly justifies Boltzmann's *Stosszahlansatz* – but *not* any quasi-deterministic particle trajectories. The concept of trajectories would approximately apply only to heavy objects that suffer mainly "pure" decoherence (with negligible recoil). "Open" quantum systems are generally described by similar phenomenological (Lindblad-type) master equations that are often *postulated* rather than being derived from realistic assumptions for a quantum environment, and then easily misunderstood as representing fundamental deviations from unitary quantum mechanics. In order to be regarded as "macroscopic" (in the sense of not being part of a thermal distribution), quasi-classical (decohered) variables have furthermore to be redundantly documented in the rest of the universe – see under "fork of causality", "consistency of documents", or "overdetermination of the past" in the first Ref. 7, for example in Footnote 1 on its page 18. Dynamically conserved information *about* such systems may nonetheless be dynamically exchanged between microscopic and macroscopic variables, that is, between negentropy and macroscopic information proper. In many situations this distinction is not very clearly defined. – In the theory of "quantum Darwinism",[30] these *classical* thermodynamic arguments are combined (and perhaps a bit confused) with the quantum concept of decoherence, which represents spreading physical entanglement, but not necessarily any spreading of (usable) information into the environment. Creation of (necessarily physical) information about a system must always lead to its decoherence, while the opposite is *not* true: even an environment in thermal equilibrium may allow further entanglement with a "system under consideration" to form. Documents which define *humanistic history* – including the history of science – obviously require even more specific correlations (which define a specific "context").

For relativistic reasons, *all* known elementary physical objects are described as quantum fields (although they are usually called "elementary particles"). The contrast between the first order in time of the Schrödinger equation and the second order of classical field equations with their negative frequencies opens the door to the concept of "anti-bosons". (For fermions this relation assumes a different form – depending on the starting point before quantization, as indicated in Footnote §.) Because of the universality of quantum fields, one may also expect a "theory of everything" to exist in the form of a unified quantum field theory. At present, though, the assumption that the fundamental arena for the universal wave function be given by the configuration space of some fundamental field(s) is no more than the most plausible attempt. On the other hand, the general framework of Schrödinger's wave function(al) or Dirac's superposition principle as a universal concept for quantum states has always been confirmed, while attempts to derive this framework from some deeper ("hidden") level have failed and are strongly restricted by various no-go theorems. Therefore, an epistemic interpretation of pure quantum states seems to be ruled out.

Among boson fields, gauge fields play a surprisingly physical role, since gauge transformations appear locally as unphysical redundancies. Their physical role is facilitated by their dynamical entanglement, which thus reveals that the redundancy holds only classically, where gauge variables appear as purely relational quantities.[18] An important question after quantization is whether gauge symmetries can be broken by a real or apparent collapse.

Unfortunately, *interacting* fields require the entanglement of such an enormous number of fundamental degrees of freedom – traditionally interpreted as "quantum fluctuations" even in time-independent states – that they cannot even approximately be treated beyond a questionable (though within its applicability very successful) perturbation theory in terms of *free* effective fields. This limitation to the quantization of coupled oscillators may also explain why the Heisenberg picture appears sufficient for QFT. Instead of consistently applying the established concepts from quantum mechanics (general superpositions) to the new variables (field amplitudes) in the form of time-dependent general field functionals, various semi-phenomenological concepts are therefore used for specific purposes – mostly for calculating scattering amplitudes between phenomenologically chosen objects that are treated as being asymptotically free. (This can be approximately valid only in exceptional situations, such as high energy laboratory experiments.) The applicability of the *S*-matrix must be quite limited, in general, since unitary dynamics is a continuous process rather than a succession of scattering events in space. It may reflect (but can hardly *explain*) properties of the "objects" to which it applies. Apparent particle properties of such asymptotic ("on shell") objects, such as their momenta, are again regarded as "real" rather than "virtual" as soon as their superpositions have irreversibly become irrelevant for local observers.

*Stable local* entanglement between different fields may be regarded as their "dressing" (similar to the entanglement between proton and electron in the bound hydrogen atom – cf. Sect. 4), while chaotic nonlocal entanglement must describe decoherence, and may thus lead to the appearance of scattering as a probabilistic rather than a unitary process. For individual field modes, such as in cavity QED, one may explicitly study their entanglement, for example that with individual atoms.

Similar semi-phenomenological methods as in QFT are also used in condensed matter physics, even when its objects of interest are non-relativistically regarded as given *N*-particle systems. They may nonetheless give rise to effective phonon fields or various kinds of "quasi-particles". The wave function for the lattice ions and their electrons, for example, is here regarded as fundamental, while the phonon field functional "emerges" – similar to Goldstone bosons in QFT. Symmetry-breaking effective ground states (such as lattices with fixed positions and orientations) and their corresponding "Fock spaces" can be understood as representing Everett branches that have become autonomous by the decoherence of their superpositions into wave packets during a condensation process.[19] Some "Fock vacua" are characterized by the number of certain particles (such as electrons in a metal) that form a *stable entanglement* in this ground state. Most familiar are pair correlations in the BCS model of superconductivity. A similar model in QFT led to the prediction of the Higgs "particle". However, only in situations described by an effective Hamiltonian that gives rise to an energy gap (defining an effective mass) can the lowest excited states approximately avoid further decoherence within their corresponding Fock space under normal conditions and low temperatures, and thus exhibit the usual phenomena of "single particle" quantum mechanics.

The BCS (pair correlation) model is also useful for understanding the meaning of Hawking or Unruh radiation,[20] which is often misinterpreted as resulting from vacuum fluctuations rather than entanglement. Since only exceptional field modes of a given space volume would obey the boundary conditions also for certain subvolumes, not even the total vacuum factorizes into *local* subvacua for the quantum field. The Hilbert space Hamiltonian, therefore, depends not only on the differential operators, but also on the chosen boundary conditions (in order to define its eigenstates), while complementary subvolumes are entangled with one another in almost all pure states of the total volume. For parallel pairs of *physical* boundaries (conducting plates), which require an infinite energy renormalization corresponding to the zero-point energies of all thereby excluded field modes, this leads to the Casimir effect as a measurable (finite) dependence on the distance between the plates. In the absence of physical boundaries, the entanglement between subvolumes may be regarded as a mutual decoherence. Non-inertial detectors, for example, define effective spacetime horizons as formal boundaries for the modes to which they couple, and thus register a thermal mixture representing Hawking or Unruh radiation in the inertial vacuum (which extends beyond these horizons). This radiation does not require any stochastic "particle creation" – there would only be stochastic registration of *apparent* particles by means of decoherence in accelerated detectors. Without this decoherence, a black hole would define a *coherent* outgoing Hawking flux, for example. The presence of particles is thus a matter of "spacetime perspective", based on the choice of non-inertial reference frames (such as Rindler frames) that are used to define their specific "plane" wave modes, while the abstract *quantum field states*, such as different "physical vacua", represent "real" states in spite of their ambiguous interpretation in terms of particles.

Within *microscopic* many-particle systems, for example in small molecules or atomic nuclei, spontaneous symmetry breaking may even lead to energy eigenstates for collective motions (such as rotations or vibrations) rather than to a real asymmetry. Since electrically neutral microscopic objects can often be assumed to be isolated from their environments, asymmetric "model ground states" (deformed nuclei or asymmetric, such as chiral, molecular configurations – similar to Fock vacua in solid states) are degenerate, and thus lead to energy bands or multiplets by means of different superpositions of all their degenerate orientations or chiralities.[21] The corresponding collective degrees of freedom are often classically visualized as describing slow ("adiabatic") motion, although classical motion would in turn require time-dependent superpositions of *different* energy eigenstates. The quantum mechanical justification of such time-dependent states, which are found for *macroscopic* objects, had to await the discovery of decoherence (here of the energy eigenstates). Since all particles in a collective superposition of different orientations are strongly entangled with one another, energy eigenstates are analogous to the bird's perspective of a quantum world, while an external observer of such an eigenstate assumes the role of a "real bird". In contrast, the whole quantum world must *contain*, and thus be entangled with, its observer, who thus gives rise to "many minds" with their asymmetric frog's perspectives (broken symmetries – cf. Sect. 4).[19] In accordance with this picture, individual particles that are part of collective rotational superpositions feel in first approximation only a *fixed* deformed potential (analogous to observing a definite measurement outcome!), as can be seen from their single-particle spectra – for example Nielson states in deformed nuclei as a variant of the nuclear shell model. (This observation originally suggested the many-minds interpretation.) In this sense, collective superpositions imitate a "multiverse" consisting of different orientations, although such quantum cosmological analogies may have delayed the acceptance of the concept of decoherence for a decade, until its "naïve" interpretation by means of the pragmatically justified reduced density matrix formalism became popular and made it acceptable to many practicing quantum physicists. In the case of a *global* symmetry, collective variables bear some similarity to gauge variables.

On a very elementary level, semi-phenomenological methods were already used for the hydrogen molecule by separately quantizing its "effective" degrees of freedom (center of mass motion, vibration, rotation and two independent electrons in the Coulomb field of adiabatically moving nuclei) rather than treating it exactly as an entangled four-body problem. Chiral molecules can at very low energies effectively be described as two-state systems, while an analogous explanation may conceivably await discovery for *all* kinds of qubits.

In QFT, the successful phenomenology of apparently fundamental fields ("elementary particles"), such as described by the Standard Model, has to be expected to form the major touchstone for any fundamental theory of the future. This may be true even though quantum chromodynamics seems to be already too complex for us to derive nuclear physics phenomena without auxiliary assumptions. This Standard Model is essentially based on linear representations of some abstract symmetry groups, whose meaning is not yet understood. The physical importance of linear representations of groups for isolated systems is just another consequence of the superposition principle. At present, however, the Standard Model does not seem to offer any convincing hints for the nature of the elusive fundamental theory.

All one may thus dare to predict is that the fundamental Hilbert space must possess a local *basis* (such as the configuration space of spatial fields and/or point-like objects) in order to allow for a definition of dynamical locality or "relativistic causality". In contrast to popular concepts of *mono*-causality, classical reality is multi-causal: in order to determine the fields at some spacetime point, one has to know them on a complete slice through its (past or future) light cone. Only because the causal connection between two *events* may then be difficult to recognize in general, did Einstein postulate the travel of "signals", which may be characterized by some identifyable structure, rather than general causal influences to be limited by the speed of light. Although *quantum superpositions* of such fields are kinematically nonlocal, and thus able to violate Bell's inequality, the *dynamical* locality defined for their basis remains valid and important for them (including interactions that describe measurements and decoherence). This relativistic causality may even prevent the formation of black hole horizons if it applies to Hawking radiation.[22] While nonlocal phase relations defining superpositions are essential for the precise value of von Neumann's conserved global ensemble entropy (zero for pure states), the dynamical transformation of information about local systems into that about nonlocal correlations or entanglement describes the increase of "physical entropy", since the latter is defined as additive and thus neglects nonlocal correlations for being thermodynamically "irrelevant".[9]

This search for the Hilbert space basis of a fundamental theory has nothing to do with that for "hidden variables", which are to explain quantum indeterminism and the wave function themselves. All novel theories that are solely based on mathematical arguments, however, have to be regarded as speculative until empirically confirmed – and even as incomplete as long as there is no general consensus about the correct interpretation of their quantization. Many quantum field theorists and mathematical physicists seem to regard their semi-phenomenological models, combined with certain methods of calculation and applied to classical field or particle concepts, as *the* quantum field theory proper. Indeed, why should one expect a conceptually consistent theory if there is no microscopic reality to be described – as assumed in the still popular Copenhagen interpretation and its variants? Therefore, most textbooks of QFT do not even *attempt* to present a consistent fundamental theory.

Our conclusion that the observed particle aspect is merely the consequence of fast decoherence processes in the detecting media does not seem to be of particular interest to many high-energy physicists, although such phenomena in their detectors are an essential part of their experiments. So some of them call the enigmatic objects of their research "wavicles", as they cannot make up their mind between particles and waves. This indifferent language represents another example of Wigner's "Balkanization of physics" (or "many words instead of many worlds" according to Max Tegmark). The wave-particle "dualism" is usually still understood in terms of *spatial* waves rather than wave functions in configuration space, although the former should by now be known to be quite insufficient in quantum theory.

## 6. Quantum Gravity and Quantum Cosmology

I cannot finish a presentation of universal quantum theory without having mentioned quantum gravity. In their linear approximation, Einstein's field equations for the metric tensor define separately oscillating spatial tensor modes, that after quantization give again rise to energy quanta $h\nu$ ("gravitons") – cf. Sect. 3. For consistency, however, the full theory has to be quantized. Its dynamical variables must then appear among the arguments of a universal wave function, and thus be entangled with all others – in a very important way, as it turns out.[23]

The Hamiltonian formalism of Einstein's nonlinear field equations, required for their "canonical" quantization, was brought into a very plausible form by Arnowitt, Deser and Misner in 1962. They demonstrated that the configuration space of gravity can be understood as consisting of the spatial geometries of all possible three-dimensional space-like hypersurfaces in spacetime. These hypersurfaces define arbitrary simultaneities that may form various foliations of spacetime, which may then be parametrized by a time coordinate *t*. This Hamiltonian form of the theory is therefore also called "geometrodynamics". Its canonical quantization leads to a (somewhat ambiguously defined) Schrödinger equation in the sense of Tomonaga for the wave functional on all these geometries – known as the *Wheeler-DeWitt equation*. This is another example demonstrating that the Hamiltonian form of a theory is not in conflict with its relativistic nature.

In contrast to the normal Schrödinger equation, the WDW equation remarkably assumes the form $H\Psi = 0$. This can also be understood as a constraint, while the Schrödinger equation itself then becomes trivial: $\partial\Psi/\partial t = 0$. The reason is that there is no classical spacetime any more to be foliated. (Each foliation would correspond to a *trajectory* through this configuration space – in conflict with quantum concepts). However, the spatial metric that occurs (besides matter variables) as an argument of the wave functional $\Psi$ would determine all proper times (defining clock times) along time-like curves which connect it classically, that is, according to the Einstein equations, with any other given spatial geometry – regardless of the choice of a foliation. Therefore, in spite of its formal timelessness, the Wheeler-DeWitt equation *does* define a physical time dependence by means of the entanglement be-tween all its variables – similar to the entanglement $\psi(u,q)$ between a quantum clock $u$ and other variables $q$ instead of a time dependence $q(u)$. Therefore, the *formal* timelessness of the WDW equation is a genuine quantum property, as it reflects the absence of trajectories. Although classical spacetimes can be represented by trajectories that can be parametrizied by one coordinate time $t$ (invariantly reparametrizable by monotoneous functions $t'(t)$), dynamical time is in general many-fingered, that is, it depends on the local progression of the spacelike hypersurfaces independently at any space point. In the case of an exactly homogenous and isotropic Friedmann cosmology, it may be represented by *one* single "finger": the expansion parameter $a$. If the wave function were regarded as a probability amplitude, it would now define probabilities *for* physical time; it is not a function *of* (some external) time any more.

It is further remarkable in this connection that, for Friedmann type universes, the Hamiltonian constraint $H\Psi = 0$ assumes a hyperbolic form in its infinite-dimensional (gauge-free) configuration space – again with $a$ or its logarithm defining a time-like variable. This property is physically quite important, as it allows for a global "initial" value problem for the wave functional – for example at $a \to 0$.[24] For increasing $a$, its solution may form a superposition of wave packets that "move" through this configuration space as a function of $a$. A drastic asymmetry of $\Psi$ with respect to a reversal of $a$ (an "intrinsic" arrow of time) might then be derivable even from symmetric boundary conditions (such as the usual integrability condition in $a$) because of the asymmetry of the Hamiltonian under this physical reversal.

Claus Kiefer could furthermore derive the time-dependent Schrödinger (Tomonaga) equation for the matter wave function under a short wave length approximation for the geometric degrees of freedom. It corresponds to a Born-Oppenheimer approximation with respect to the inverse Planck mass (see Kiefer's Ch. 4 in Joos et al. of Ref. 7, or his Sect. 5.4 of Ref. 23). This result demonstrates that the Wheeler-DeWitt equation can only describe a whole Everett multiverse, since each trajectory in the configuration space of spatial geometries may define its own classical spacetime. Narrow wave packets for the spatial geometry approximately propagating along such trajectories are decohered from one another by the matter variables (which thereby serve as an "environment"). This is analogous to the decoherence of atomic nuclei in large molecules by collisions with external particles – the reason why they appear to move on quasi-classical trajectories according to the frog's perspective of a human observer. In cosmology, decoherence (that is, uncontrollable entanglement rather than the often mentioned "quantum fluctuations") is also important for the origin of "classical" structure in the early universe during the onset of inflation.[25]

If one also allows for "landscapes" (Tegmark's Level 2 of multiverses[26]), which are assumed to exist in several hypothetical cosmologies that lead to a drastically inhomogeneous universe on the very large scale, the "selection" (by chance – not by free will) of a subjective observer with his epistemologically important frog's perspective (cf. Sect. 4) may be roughly characterized by a hierarchy of five not necessarily independent steps: (1) the selection (in the sense of Everett or Tegmark's Level 3 – usually regarded as a quantum measurement) of an individual landscape from their superposition that must be part of a global quantum state, (2) the selection of a particular region in this three or higher dimensional landscape (a causally separate "world" that may even be characterized by specific values of certain "constants of nature" – Level 2), (3) the selection of a quasi-classical spacetime from the Wheeler-DeWitt wave function by means of decoherence as indicated above (Level 3 again), (4) the selection of one individual complex organism from all those that may exist in this "world", including some "moment of awareness" for it (giving rise to an approximate localization of this observer in space and time: a subjective "here-and-now" – thus including Tegmark's Level 1), and (5) the selection of one of his/her/its "versions" that must have been created by further Everett branching based on the decoherence of matter variables according to Sect. 4 (Level 3). Therefore, every conceivable subjective observer who is part of the universe requires an extreme "individualization" (multifold localization) *in* the complete quantum world, and hence *of* his "Everett world" (his frog's perspective). It seems to be required in order to define appropriate IISs (integrated-information systems), IGUSs (information gaining and utilizing systems), or whatever you call such parts of the universal wave function that may form the physical basis for conscious beings, and, therefore, for an *observable* universe. New *physical* laws may not be required for this purpose.

Each of these probabilistic steps may create its own unpredictable initial conditions characterizing the further evolution of the resulting individual "worlds". Most properties characterizing our observed one can thus not be derived from any physical theory; they have to be empirically determined as part of an answer to the question: *Where* do *we* happen to live in objective "configuration" space? This unpredictability, including that of certain "constants of nature", is by no means specific for a multiverse (as some critics of this concept argue). It would similarly apply to any kind of stochastic dynamics (such as collapse theories), or whenever statistical fluctuations are relevant during the early cosmic evolution. Only step 4 can *not* be objectivized in the usual sense, namely with respect to different observers in the same quasi-classical world. Some of these steps may require an application of the weak anthropic principle in some sense (although I would not recommend to *rely* on it by playing

"Russian quantum roulette"!). Although each *individual* outcome can only have very small probability because of their large number, the actually observed ones should not be exceptionally improbable among them. This is still a strong condition, which may suffice to explain why observed frequencies of measurement results are in accord with Born's rule (Sect. 4). Entropy may *decrease* during most of these steps (depending on its precise definition).[6,9,27]

Let me add for completeness that Tegmark's Level 1 and 2 multiverses are classical concepts, and thus unrelated to Everett's branches, as they merely refer to separate regions in conventional space rather than branches in "configuration" space. It appears somewhat pretentious here to speak of "parallel worlds" or "multiverses"; these names were originally invented for Everett branches, and are here simply misused. This may perhaps be explained by the fact that many cosmologists had never accepted the role of entangled superpositions as part of quantum reality, and therefore prefer to replace them explicitly or tacitly by *statistical* correlations characterizing a collapse mechanism, for example. In this case, different outcomes could be realized only at different places in a spatially sufficiently large universe, while – in contrast – different Everett "worlds" exist (in "configuration" space) even for a closed universe of traditional size. (However, even different kinds and sizes of universes may formally exist in one superposition if the superposition principle holds for them, too.)

While landscapes with regions of different fundamental properties would be quite plausible in a spatially unbounded or very large universe without making use of Everett (similar to locally varying order parameters resulting from symmetry breaking phase transitions in solid state physics[19]), almost *identical* local situations occurring by chance somewhere in an infinite quasi-homogeneous world (Level 1) may be regarded as something between trivial (entirely irrelevant for us) and ill-defined. Although the double exponentials which are required to describe the expected distances from such statistical Doppelgängers can easily be *formulated*, an extrapolation of observed properties (such as an approximately flat quasi-classical space) from the observable universe with its size of $10^{10}$ *ly* to something like $10^{10^{10^{100}}}$ *ly* appears at least risky. Statistical estimates of probabilities would in any case apply only to chance fluctuations (such as "Boltzmann brains"), but not to situations resulting from evolution. If their probabilities, are calculated by means of some *physical* (that is, additive) entropy, they would completely neglect the existence of "consistent documents" (often regarded as an "overdetermination of the past" – see Footnote ** above and Sect. 3.5 of the first Ref. 9), while *unstructured* initial conditions (such as the initial homogeneity of a gravitating universe) represent even lower entropy values – in spite of their "plausibility".

The role of Tegmark's (as yet unmentioned) Level 4 universes is entirely questionable, since mathematics, while providing extremely useful conceptual tools for physics because of its analytical (tautological) nature and, therefore, the undeniable *formal truth* of its theorems, cannot by itself warrant the *applicability* of its formal concepts to the empirical world. Only if, and insofar as, such kinematical concepts have been empirically verified to be consistently applicable in a certain context, can we consider them as candidates for a description of "reality". (This seems to be a point that many mathematicians working in theoretical physics and cosmology have problems to understand, since they are used to defining their concepts just for convenience.) Different mathematical frameworks can therefore not be regarded as indicating the existence of corresponding different physical "worlds" or different parts of one multiworld. While Everett's "many worlds" (just as all scientific cosmology) result from hypothetical extrapolation of the observed world by means of empirical laws, there are no reasons supporting the physical existence of Level 4 worlds. The *mathematical* concept of "existence", for example, means no more than the absence of logical inconsistencies, that is, a necessary (hence important[28]) but not a sufficient condition for being "realized" in Nature.

## 7. Conclusions

These remarks about quantum gravity and quantum cosmology may bring the strange story of particles and waves to a preliminary end. While the particle concept was recognized as a delusion, the observed wave aspects of microscopic objects can be understood only as part of a universal wave function in a very-high-dimensional (if not infinite-dimensional) "configuration" space. If this wave function describes reality completely, no *local* properties can generically "exist" (although they may be *observed*). Only if one insisted that reality *must* be defined in space and time would this very concept of reality (which is essentially equivalent to the existence of a consistent description of Nature) have to be abandoned. In spite of Bell's theorem, the prejudice that reality has to be kinematically local still seems to represent the major hurdle for many physicists to accept a universal wave function as a *physical* object.

The *observable* quantum world that defines our frog's perspective (that is, the objectivized relative state with respect to "individual minds" – see Sect. 4) is thus no more than a tiny component of this global wave function. The latter, representing the "bird's view", may be regarded as *the true hidden reality* behind the phenomena, since it is required from the observer's point of view only for reasons of dynamical consistency. This may appear similar to the assumed reality of spacetime in classical GR in spite of the specific reference frames to which observers are bound. Ongoing branching of the relative state into autonomous sub-branches of the wave function by means of decoherence then mimics a collapse process, and in this way solves the measurement problem in terms of many (branching) minds. The full Wheeler-DeWitt wave function, for example, seems to be meaningful only from the bird's perspective (see Sect. 6).

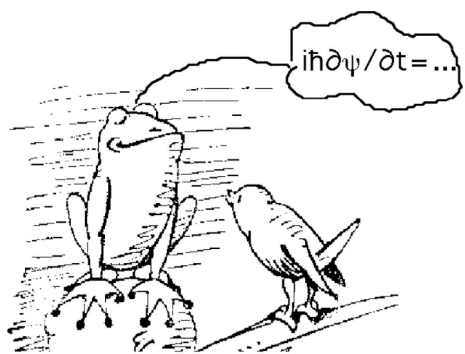

**Fig. 2:** A frog's bird's perspective

Matrix mechanics with its formal concept of "observables" thus turns out to be only an effective probabilistic description in terms of not consistently applicable (hence mutually "complementary") particle or other traditional concepts, which may in certain situations approximately apply to the observed world (our branch). Nonetheless, many physicists are still busy constructing absurdities, paradoxes, or no-go theorems in terms of such traditional concepts in order to demonstrate the "weirdness" of quantum theory. This includes the infamous "information loss paradox" for black holes, which are predicted to disappear by means of Hawking radiation (a quantum phenomenon), but are also assumed to possess a classical event horizon of some kind.[22] Even Alice and Bob are classical concepts that have to be justified by means of decoherence, that is, nonlocal entanglement with their environment. "Quantum Bayesianism", presently much *en vogue*, replaces the whole physical world by a black box, representing an abstract concept of "information" about inconsistent classical variables, and assumed to be available to vaguely defined "agents" rather than to observers who may be consistent parts of the physical world to be described. In contrast to Everett's Many Worlds, for example, such a "non-theory" can never be falsified (it is "not even wrong").

While effective concepts like particles and spatial fields remain important for our every-day life, including that in physics laboratories, their limited validity must deeply affect a consistent world model (cosmology, in particular). It is always amazing to observe how the love affair of mathematical physicists and general relativists with their various classical field theories often prevents them from accepting, or even from sufficiently understanding, non-local quantum states that are well-known from elementary quantum mechanics. Quantum effects are then often belittled as mere "anomalies" of their theories. Some mathematical physicists are even trying to "explain" non-local quantum entanglement by means of speculative "worm holes" in space – apparently an attempt to save their belief in some local reality.

We have to accept, however, that the precise structure of a local Hilbert space *basis*, which is often assumed to be given by the configuration space of some fundamental fields, remains elusive. Because of the unavoidable entanglement of all variables, one cannot expect the *effective* quantum fields, which are said to describe "elementary particles", to be related to these elusive fundamental variables in a simple way. This conclusion puts in doubt much of the traditional approach to QFT, which is based on concepts of renormalization and "dressing". There are indeed excellent arguments why even emergent ("effective") or quasi-classical fields may be mathematically elegant – thus giving rise to the impression of their fundamental nature. Novel mathematical concepts might nonetheless be required for finding the elusive ultimate theory, but their applicability to physics would have to be demonstrated empirically, and can thus never be confirmed to be *exactly* valid. This may severely limit the physical value of many "abstract" (non-intuitive) mathematical theorems. Just think of Einstein's words "*Insofern sich die Sätze der Mathematik auf die Wirklichkeit beziehen, sind sie nicht sicher, und insofern sie sicher sind, beziehen sie sich nicht auf die Wirklichkeit*", or Feynman's remark regarding early attempts to quantize gravity:[3] "Don't be so rigorous or you will not succeed." Fundamental physical laws and concepts have so far mostly turned out to be mathematically relatively simple, while their applications may be highly complex. This fact may explain why mathematicians have dominated theoretical physics mostly *after* completion of a new fundamental theory (such as Newton's and even more so Einstein's – but *not* yet for quantum theory!), or at times of stagnation, when mere reformulations or unconfirmed formal speculations (such as strings at the time of this writing) are often celebrated as new physics.